\pdfoutput=1

\documentclass[12pt]{article}
\usepackage{epsfig}

\def\beq{\begin{equation}}
\def\eeq#1{\label{#1}\end{equation}}
\def\eeqn{\end{equation}}

\def\beqa{\begin{eqnarray}}
\def\eeqa#1{\label{#1}\end{eqnarray}}
\def\eeqan{\end{eqnarray}}

\let\bar=\overbar

\def\Dslash{\not{\hbox{\kern-4pt $D$}}}
\def\dslash{\not{\hbox{\kern-2pt $\del$}}}

\def\msb{{\bar{\ssstyle M \kern -1pt S}}}

\def\Title#1{\begin{center} {\Large {\bf #1} } \end{center}}

\begin{document}

\Title{Threshold Studies of Charm Mixing and Strong Phases with CLEO-c}

\bigskip\bigskip


\begin{raggedright}  

{\it Werner Sun (for the CLEO Collaboration)\\
Physical Sciences Building\\
Cornell University\\
Ithaca, NY 14853\\
USA}
\bigskip\bigskip
\end{raggedright}

\section{Introduction}

Charm mixing in the Standard Model is described by the parameters
$x \equiv 2(M_2 - M_1)/(\Gamma_2 + \Gamma_1)$ and
$y \equiv (\Gamma_2 - \Gamma_1)/(\Gamma_2 + \Gamma_1)$,
where $M_{1,2}$ and $\Gamma_{1,2}$ are the masses and widths, respectively,
of the neutral $D$ meson $CP$ eigenstates, $D_1$ ($CP$-odd) and
$D_2$ ($CP$-even):
$|D_{1,2}\rangle \equiv (|D^0\rangle \pm |\bar D^0\rangle)/\sqrt{2}$.
Recently, several experiments have directly probed $x$ and
$y$~\cite{ycpBelle1,ycpBABAR1,ycpBABAR2,ycpBelle2,ycpLHCb,kspipiBelle,kspipiBABAR},
as well as the ``rotated'' parameters
$y'\equiv y\cos\delta - x\sin\delta$ and
$x'\equiv y\sin\delta + x\cos\delta$~\cite{kpiBelle,kpiBABAR,kpiCDF},
where
$-\delta$ is the relative phase between the doubly
Cabibbo-suppressed $D^0\to K^+\pi^-$ amplitude and the corresponding
Cabibbo-favored $\bar D^0\to K^+\pi^-$ amplitude:
$\langle K^+\pi^-|D^0\rangle / \langle K^+\pi^-|\bar D^0\rangle\equiv r e^{-i\delta}$.
We adopt a convention in which $\delta$ corresponds to a strong
phase, which vanishes in the SU(3) limit~\cite{Gronau:2001nr}.
In this article, we update an analysis~\cite{tqca1} that gave the first 
direct determination of
$\cos\delta$ using correlated production of $D^0$ and $\bar D^0$ mesons 
in $e^+e^-$ collisions produced at the Cornell Electron Storage Ring and
collected with the CLEO-c detector.
Here, we also present a first measurement of $\sin\delta$.

At the $\psi(3770)$ resonance,
the $D^0\bar D^0$ system is produced in an overall charge conjugation
eigenstate with $C=-1$. As a result of this quantum coherence, the
exclusive
$D^0\bar D^0$ decay rate to a final state $\{i,j\}$, where $i$ and
$j$ label the final states of the two $D$ mesons, is given by the
square of the antisymmetric amplitude
\beq
\Gamma(i,j)\propto {\cal M}^2_{ij} \equiv \left|\langle i|D^0\rangle\langle j|\bar D^0\rangle - \langle i|\bar D^0\rangle\langle j|D^0\rangle \right|^2 + {\cal O}(x^2, y^2),
\eeqn
where the ${\cal O}(x^2, y^2)$ term represents a mixed amplitude.
These exclusive rates depend on the mixing parameters and amplitude
ratios defined above:
\beqa
\Gamma(i,\bar\jmath) = \Gamma(\bar\imath, j) &\propto&
  1 + r_i^2 r_j^2 - 2 r_i \cos\delta_i \ r_j \cos\delta_j - 2 r_i \sin\delta_i \ r_j \sin\delta_j \\
\label{eq:ratesDCS}
\Gamma(i,j) = \Gamma(\bar\imath, \bar\jmath) &\propto&
  r_i^2 + r_j^2 - 2 r_i\cos\delta_i \ r_j\cos\delta_j + 2 r_i\sin\delta_i \ r_j\sin\delta_j.
\eeqan
Inclusive rates are obtained by summing over exclusive rates and are
trivially related to the branching fractions (${\cal B}$) in uncorrelated
decay:
\beq
\Gamma(i,X) = \sum_j\left[ \Gamma(i,j) + \Gamma(i,\bar\jmath) \right ] =
{\cal B}_i + {\cal B}_{\bar\imath} \propto
1 + 2 y r_i \cos\delta_i + r_i^2.
\eeqn
By comparing correlated and uncorrelated decay rates, we extract
simultaneously the charm mixing and strong phase parameters, as
well as the number of $D^0\bar D^0$ pairs produced (${\cal N}$)
and the branching fractions of the reconstructed
$D^0$ final states.

Compared to our previous analysis~\cite{tqca1}, this update uses a
dataset three times larger (818 ${\rm pb}^{-1}$), and we reconstruct
more $CP$ eigenstates,
as well as semimuonic $D^0$ decays and modes that provide
sensitivity to $\sin\delta$ and $r$.

\section{Experimental Technique}

We reconstruct the $D$ meson final states listed in
Table~\ref{tab:finalStates}. 
Following Ref.~\cite{kspipi}, the $K^0_S\pi^+\pi^-$
Dalitz plot is divided into eight bins according to the strong phase of
the decay amplitude.
Because $CP$ eigenstates and
semileptonic final states have known values of $r_i$ and $\delta_i$,
they act as reference points for determining $y$ and the strong
phase $\delta$ in the $K\pi$ final state.
Inclusive rates are derived from yields of single tags (ST),
or individually reconstructed $D^0$ or $\bar D^0$ candidates.
For exclusive rates, we reconstruct $D^0\bar D^0$ pairs or
double tags (DT).

\begin{table}[b]
\begin{center}
\begin{tabular}{ccccc}
\hline\hline
Type & Reconstruction & Final States & $r_i$ & $\delta_i$ \\
\hline
Mixed-$CP$ & Full & $K^-\pi^+$, $K^+\pi^-$, $K^0_S\pi^+\pi^-$ &
\multicolumn{2}{c}{From fit} \\
$CP+$ & Full & $K^+K^-$, $\pi^+\pi^-$, $K^0_S\pi^0\pi^0$ & 1 & $\pi$ \\
$CP+$ & Partial & $K^0_L\pi^0$, $K^0_L\eta$, $K^0_L\omega$ & 1 & $\pi$ \\
$CP-$ & Full & $K^0_S\pi^0$, $K^0_S\eta$, $K^0_S\omega$ & 1 & 0 \\
$CP-$ & Partial & $K^0_L\pi^0\pi^0$ & 1 & 0 \\
Semileptonic & Partial & $K^- \{e^+,\mu^+\}\nu_{e,\mu}$,
$K^+ \{e^-,\mu^-\}\bar\nu_{e,\mu}$ & 0 & ---\\
\hline\hline
\end{tabular}
\caption{$D$ final states reconstructed in this analysis.}
\label{tab:finalStates}
\end{center}
\end{table}

Fully reconstructed modes are identified using two kinematic variables:
the beam-constrained candidate mass
$M \equiv \sqrt{ E_0^2 / c^4 - {\mathbf p}_D^2 / c^2 }$ and the
energy difference
$\Delta E \equiv E_D - E_0$,
where ${\mathbf p}_D$ and $E_D$ are the total momentum and energy of the $D$
candidate, respectively, and $E_0$ is the beam energy.
We measure ST yields for all fully reconstructed modes except
$K^0_S\pi^+\pi^-$.
Decays with $K^0_L$ mesons and neutrinos, which we do not detect directly,
are identified with a partial reconstruction technique,
where the presence of the undetected particle is inferred via
conservation of energy and momentum.
We form DT candidates from all combinations of modes in
Table~\ref{tab:finalStates}, both Cabibbo-favored and
Cabibbo-suppressed, with at most one missing particle.
In addition, we reconstruct $\{Ke\nu_e, K^0_L\pi^0\}$, which has two
missing particles, using the technique described in
Refs.~\cite{Brower:1997be, Hokuue:2006nr}.

\section{Results}

We combine 261 yield measurements, along with estimates of efficiencies
and background contributions, in a least squares fit that determines 51
free parameters, including the five charm mixing and $D\to K\pi$
amplitude parameters listed in Table~\ref{tab:results}.
The other parameters, for which the fit results are not shown,
 are ${\cal N}$, amplitude ratios and phases for the 8 phase bins in
$K^0_S\pi^+\pi^-$, and 21 branching fractions. 
The fit includes both statistical and systematic uncertainties on the
input measurements. We separate the statistical and systematic
uncertainties in Table~\ref{tab:results} by repeating the fit with only
statistical uncertainties on the inputs.
We perform one fit with no external inputs
(Standard Fit) and another including the external measurements of
$y$, $x$, $r^2$, $y'$, and $x'^2$
compiled in Refs.~\cite{hfag, hfag2010} (Extended Fit).
In the Standard Fit, there is a sign ambiguity for $\sin\delta$, which
is resolved in the Extended Fit.
Because we treat uncertainties on external measurements as systematic
uncertainties, when a fit parameter is directly constrained by an
external measurement, we quote only one uncertainty in Table~\ref{tab:results};
in these cases, the statistical uncertainty effectively vanishes.

\begin{table}[b]
\begin{center}
\begin{tabular}{lcc}
\hline\hline
Parameter & ~~~~~~~Standard Fit~~~~~~~ & ~~~~~~~Extended Fit~~~~~~~ \\
\hline
$y$ (\%)
  & $4.2\pm 2.0\pm 1.0$ & $0.636\pm 0.114$ \\
$r^2$ $(\%)$
  & $0.533\pm 0.107\pm 0.045$ & $0.333\pm 0.008$ \\
$\cos\delta$
  & $0.81^{+0.22+0.07}_{-0.18-0.05}$ & $1.15^{+0.19+0.00}_{-0.17-0.08}$ \\
$\sin\delta$
  & $-0.01\pm 0.41 \pm 0.04$ & $0.56^{+0.32+0.21}_{-0.31-0.20}$ \\
$x^2$ $(\%)$
  & $0.06\pm 0.23\pm 0.11$ & $0.0022\pm 0.0023$ \\
\hline\hline
\end{tabular}
\caption{Results from the Standard Fit and the Extended Fit for
all parameters except branching fractions.
Uncertainties are statistical and systematic, respectively.
In the Extended Fit, we quote only one uncertainty for $y$, $r^2$, and $x^2$,
which are directly constrained by an external measurement.}
\label{tab:results}
\end{center}
\end{table}

In the Standard Fit, the statistical uncertainties on $y$ and $\cos\delta$
are roughly three times smaller than in our previous
analysis~\cite{tqca1}. The Extended Fit demonstrates that our measurements
of $\cos\delta$ and $\sin\delta$ can be used to improve the
uncertainty on $y$ (by combining $y$ and $y'$ from other experiments)
by approximately 10\%, compared to the current world average
found in Ref.~\cite{hfag2010}: $y = 0.79\pm 0.13$.

\begin{figure}[htb]
\begin{center}
\epsfig{file=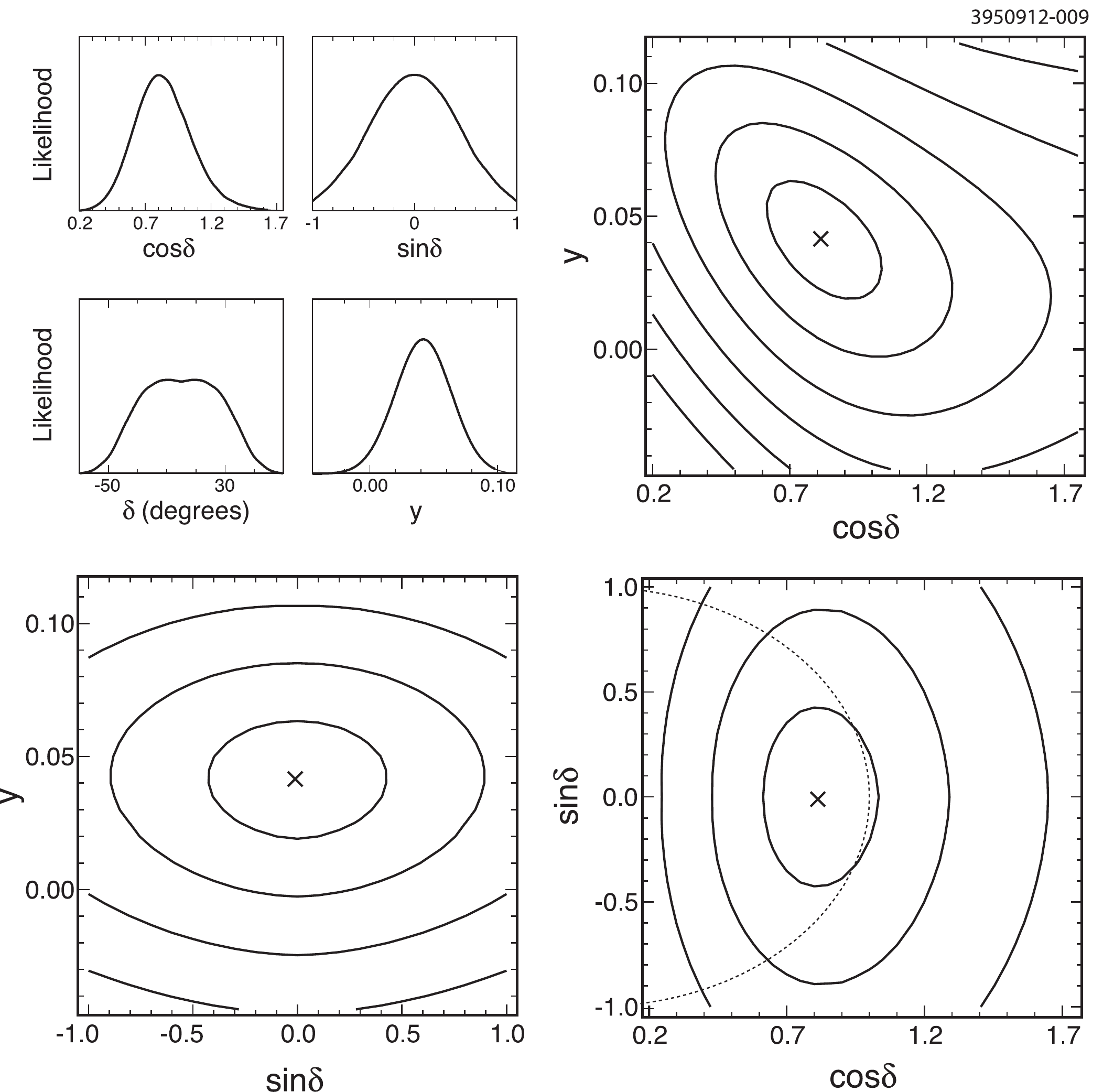,width=0.49\linewidth}
\epsfig{file=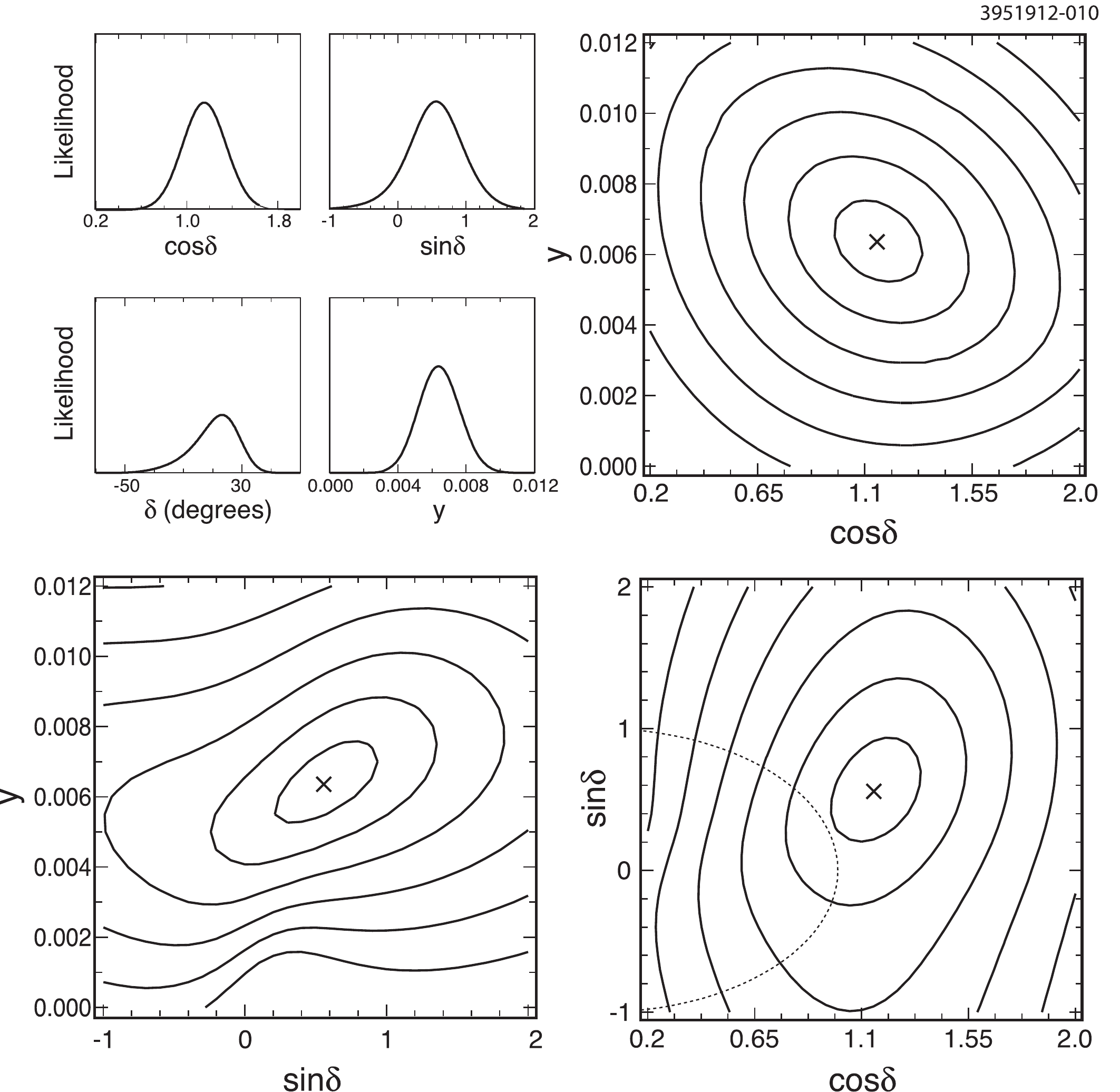,width=0.49\linewidth}
\caption{Likelihoods for the Standard Fit (left) and Extended Fit (right)
including both statistical and systematic
uncertainties for $\cos\delta$, $\sin\delta$, $\delta$, and $y$. The
two-dimensional likelihoods are shown as solid contours in increments of
$1\sigma$, where $\sigma=\sqrt{\Delta\chi^2}$.
The dashed contour marks the physical boundary, where
$\cos^2\delta+\sin^2\delta = 1$.}
\label{fig:contours}
\end{center}
\end{figure}

Asymmetric uncertainties are determined from the posterior probability
distribution functions (PDFs) shown in Fig.~\ref{fig:contours}.
These curves are
obtained by re-minimizing the $\chi^2$ at each point and computing
${\cal L} = e^{-(\chi^2-\chi^2_{\rm min})/2}$.
We construct the PDFs for $\delta$ by scanning $\cos\delta$ and
$\sin\delta$ under the constraint $\cos^2\delta+\sin^2\delta = 1$, which
result in implied values for $\delta$ of
$|\delta| = (10^{+28+13}_{-53-0})^\circ$ for the Standard Fit and
$\delta = (18^{+11}_{-17})^\circ$ for the Extended Fit.

\section{Summary}

We present an updated analysis of charm mixing and the $D\to K\pi$
strong phase using quantum correlations in $D^0\bar D^0$ decays
at the $\psi(3770)$ resonance. These results are based on the full
CLEO-c dataset, and they make a significant contribution to the
world averages of mixing parameters.

\end{document}